\documentclass[a4paper,11pt]{article}
\usepackage{jcappub}
\usepackage[utf8]{inputenc}
\usepackage{amsmath}
\usepackage{amssymb}
\usepackage{xcolor}
\usepackage[normalem]{ulem}
\newcommand{\CLASS}{\texttt{CLASS}}
\newcommand{\MontePython}{\texttt{MontePython}}

\title{Cosmological constraints on TeV-scale dark matter subcomponents decaying \mbox{between recombination and reionisation}}
\author[a,b]{Markus R. Mosbech,}
\emailAdd{mosbech@physik.rwth-aachen.de}
\author[a,c]{Cristina Benso}
\emailAdd{cristinabenso92@gmail.com}
\author[a,c]{and Felix Kahlhoefer}
\emailAdd{kahlhoefer@kit.edu}

\affiliation[a]{Institute for Theoretical Particle Physics (TTP), Karlsruhe Institute of Technology (KIT), 76128 Karlsruhe, Germany}
\affiliation[b]{Institute for Theoretical Particle Physics and Cosmology (TTK), RWTH Aachen University, D-52056 Aachen, Germany}
\affiliation[c]{Institute for Astroparticle Physics (IAP) Karlsruhe Institute of Technology (KIT), Hermann-von-Helmholtz-Platz 1, 76344 Eggenstein-Leopoldshafen, Germany}

\abstract{The Dark Ages and the Cosmic Dawn are an untapped well of information about the particle physics properties of dark matter, which may become accessible with future radio telescopes able to probe the 21-cm signal from atomic hydrogen. In this work we study the impact on cosmological observables of a dark matter subcomponent composed of TeV-scale particles that decay into electrons, photons or neutrinos with a lifetime shorter than the age of the universe. We re-evaluate constraints from the Cosmic Microwave Background (CMB) on these scenarios using the most recent data sets and estimate the sensitivity of future detections of the global 21-cm signal. Our main result is that the latter is potentially more sensitive to the effects of decaying dark matter with a lifetime $\tau \gtrsim 10^{15} \, \mathrm{s}$. This effect is strongest for the case of decays into neutrinos due to the different spectral distribution of the injected electromagnetic energy. For DM masses well above the TeV-scale, these differences become less pronounced and the sensitivity of both the CMB and the 21-cm signal depend primarily on the total amount of injected electromagnetic energy.}


\begin{document}
\begin{flushright}
        {\tt TTP26-013}\\
        {\tt TTK-26-08 }
\end{flushright}	
\maketitle

\section{Introduction}

All cosmological data is consistent with the assumption that dark matter (DM) is a perfectly cold and collisionless fluid with conserved comoving number density. While this assumption is plausible from the point of view of typical particle physics models of DM, deviations from the simple picture are expected to emerge at some level, because DM particles are generally predicted to experience scattering, annihilation and/or decay processes. This prediction motivates analyses searching for deviations from the simplest assumptions using data sets probing different periods of the early universe from before recombination~\cite{Acharya:2019uba,Lucca:2019rxf,Bolliet:2020ofj,Balazs:2022tjl,Liu:2023fgu,Liu:2023nct} until after reionisation~\cite{Liu:2020wqz,Capozzi:2023xie,Xu:2024uas,Lopez-Honorez:2026bzj}. The main conclusion from such analyses is that many effects that would reveal the particle nature of DM are tightly constrained by data~\cite{Poulin:2016anj,Slatyer:2016qyl}. This is true even for processes that happen during the Dark Ages and Cosmic Dawn, between recombination and reionisation, an epoch that we have not yet been able to directly observe~\cite{Chen:2003gz,Slatyer:2012yq,Slatyer:2015jla,Slatyer:2015kla,Liu:2016cnk}. Most notably, only a tiny fraction of DM -- at the level of $10^{-9}$ or below --  is allowed to convert into energetic electromagnetic radiation between recombination and reionisation. The reason is that even small changes in the free electron fraction potentially affect the temperature anisotropies of the Cosmic Microwave Background (CMB)~\cite{Adams:1998nr,Padmanabhan:2005es,Slatyer:2009yq,Lopez-Honorez:2013cua}, which have been measured with great precision by satellites~\cite{Planck:2019nip} and ground-based telescopes~\cite{SPT-3G:2024atg,ACT:2023kun}.

A potentially even more sensitive probe of the Dark Ages and Cosmic Dawn stems from observations of the 21 cm line arising from the hyperfine transition of neutral hydrogen~\cite{Pritchard:2010pa,Facchinetti:2023slb,Sun:2023acy}. The observed intensity of this line depends on the so-called spin temperature of the hydrogen gas, which can differ from the temperature of CMB photons, leading to an absorption of CMB photons~\cite{2012RPPh...75h6901P,Furlanetto:2006jb}. This absorption feature is expected to be particularly striking during the so-called Cosmic Dawn, corresponding to the formation of the first stars at redshift 10--20. Exotic energy injection may heat the intergalactic medium and thereby modify the absorption feature in observable ways, both in terms of the sky-averaged (monopole) signal~\cite{Monsalve:2016xbk,Singh:2017syr,Spinelli:2022xra,deLeraAcedo:2022kiu} and in the power spectrum inferred from its fluctuations~\cite{Tingay:2012ps,Pober:2013ig,LOFAR:2013jil,DeBoer:2016tnn}. In the present work we focus on the sky-averaged signal, which in spite of large astrophysical uncertainties provides a sensitive probe of the detailed properties of DM~\cite{Valdes:2012zv,Evoli:2014pva,Lopez-Honorez:2016sur,DAmico:2018sxd,Hiroshima:2021bxn,Cheung:2018vww,Agius:2025xbj,Sun:2025ksr}. Rather than studying the signal claimed by EDGES~\cite{Bowman:2018yin} and disputed by SARAS~\cite{Singh:2021mxo}, we study the potential sensitivity of future measurements.

The case that DM decays into electromagnetic particles with a lifetime much longer than the age of the universe has been studied in great detail~\cite{Diamanti:2013bia,Xu:2024vdn} and it has been shown that in many cases observations of the 21-cm line may provide stronger bounds on the lifetime than CMB observations~\cite{Clark:2018ghm,Mitridate:2018iag,Liu:2018uzy,Facchinetti:2023slb}. More recently, also the case of DM subcomponent that decays before reionisation has been studied in detail~\cite{Cima:2025zmc}. In this case, the exotic energy injection is relevant in a more narrow redshift range, which may change the relative strength of CMB constraints and 21-cm sensitivity projections depending on the assumed lifetime. These previous studies have focused on DM masses below the TeV scale decaying into photons or electron-positron pairs. In the present work, we extend these studies by going beyond the TeV scale and considering also decays into neutrinos. While decays into neutrinos inject a negligible amount of electromagnetic energy for sub-TeV DM particles, for heavier particles the final-state radiation of electroweak gauge bosons becomes more and more relevant. For example, a 10 TeV particle decaying into a pair of neutrinos injects almost 10\% of its rest mass in electromagnetic channels~\cite{Cirelli:2010xx}.

It was argued in ref.~\cite{Hambye:2021moy} that bounds on DM decays into neutrinos can be obtained simply by rescaling the fraction of decaying DM with the fraction of electromagnetic energy produced in the decay. In the present work we revisit this approximation and show that -- while it works reasonably well for CMB constraints -- it does not give reliable results for projecting the sensitivity of 21-cm intensity measurements. The reason is that in the relevant redshift range, energy deposition is not always instantaneous and the mean free path of the injected particles may depend sensitively on their energy. Only for the highest masses (10 TeV) and shortest lifetimes ($3 \cdot 10^{14} \, \mathrm{s}$) considered our analysis do we recover the simple scaling with the fraction of injected electromagnetic energy. For these points, however, the projected sensitivity of 21-cm observations is not competitive with existing bounds from the CMB in all but the most optimistic scenarios. For longer lifetimes around $10^{16} \, \mathrm{s}$, on the other hand, we find that the case of DM decays into neutrinos offers a particularly attractive target in the sense that even under pessimistic assumptions it will be possible to improve upon CMB constraints with future 21-cm observations.

The remainder of this work is structured as follows. In section~\ref{sec:model} we motivate the model that we study and introduce the set-up for our subsequent analysis. Constraints from the CMB and sensitivity estimates for the 21-cm signal are discussed in sections~\ref{sec:CMB} and~\ref{sec:21cm}, respectively. We then present our results in section~\ref{sec:results} before concluding in section~\ref{sec:conclusion}.

\section{Model set-up}
\label{sec:model}

Our analysis is based on the hypothesis that the total amount of DM present in the universe before recombination consists of several different components, one of which decays between recombination and today. This sub-component is parametrised by the fraction of the total DM energy density that it constitutes initially $f_\text{decay}$ and the lifetime $\tau_\chi$. In other words, the energy density of this sub-component is given by
\begin{equation}
    \rho_\chi(t) = f_\text{decay} \rho_\text{DM,0} (1+z)^3 e^{-t / \tau_\chi} \; ,
\end{equation}
where $\rho_\text{DM,0} = \Omega_\text{DM} \rho_\text{crit,0}$ is the present-day energy density of DM. Here we have assumed that $f_\text{decay}$ is so small that the effect of the decays on the total energy density of DM is completely negligible. Indeed, we will be interested in the range $10^{-11} < f_\text{decay} < 10^{-7}$, where this assumption is fully justified. We restrict our analysis to this range, which represents a wide band around existing limits.

Nevertheless, the decays of the subcomponent can have an observable effect on the evolution of the universe, if they inject electromagnetic energy into the intergalactic medium (IGM). To study this injection in detail, we introduce as a third parameter the mass $M_\chi$ of the decaying particle and consider annihlations into photon pairs ($\gamma\gamma$), electron-positron pairs ($ee$) and pairs of electron neutrinos ($\nu\nu$). In the first two cases, nearly all of the energy of the decaying particles is injected in the form of electromagnetic energy. In the case of decays into neutrinos, most of the energy goes directly into the neutrino pair, which have negligible probability of interacting with the IGM, such that they do not affect the subsequent evolution. However, for $M_\chi$ greater than a few hundred GeV, there is a non-negligible probability that the highly-energetic neutrinos radiate electroweak gauge bosons which decay into electrically charged particles. As a result, even DM particles that decay into neutrinos inject a fraction $\zeta_\text{em}(M_\chi)$ of their energy density in the form of electromagnetic energy.

\begin{figure}
    \centering
    \includegraphics[width=0.6\linewidth]{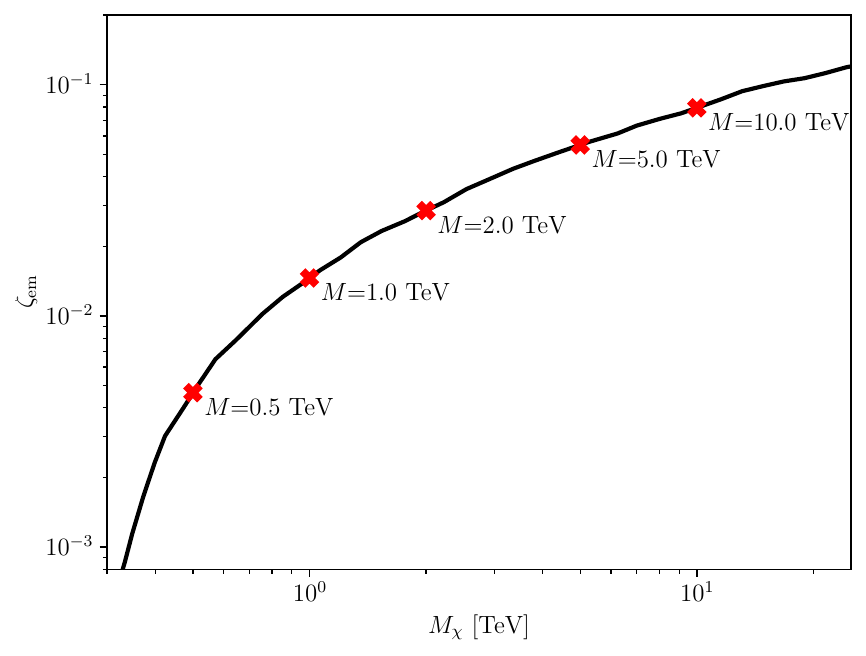}
    \caption{Efficiency factor of conversion to electromagnetic energy for DM decay into neutrinos, following ref.~\cite{Hambye:2021moy}.}
    \label{fig:zeta}
\end{figure}

The function $\zeta_\text{em}$ is shown in figure~\ref{fig:zeta}. Here we use the PPPC injection spectra~\cite{Cirelli:2010xx}, which agree reasonably well with more recent estimates~\cite{Bauer:2020jay,Hambye:2021moy} and have the advantage that they are readily available in the public code \texttt{DarkHistory}~\cite{Liu:2019bbm,Sun:2022djj}, which we use for our analysis (see below). In the present study we focus on $M_\chi > 500 \, \mathrm{GeV}$, for which $\zeta_\text{em} > 0.5\%$, noting that the case of DM subcomponents with $M_\chi < 1 \, \mathrm{TeV}$ decaying into photons and electrons has been discussed in detail in several recent studies~\cite{Cima:2025zmc,Lopez-Honorez:2026bzj}. An upper bound of $M_\chi < 10 \, \mathrm{TeV}$ is imposed by the pre-tabulated functions implemented in \texttt{DarkHistory}. However, a key result of our analysis will be a prescription for how to extrapolate our results to even heavier DM masses.\footnote{Spectra for higher energies have been computed in e.g. Refs. \cite{Bauer:2020jay,Arina:2023eic}, but they are not implemented in \texttt{DarkHistory} at this point.}

The injection rate of electromagnetic energy is therefore given by~\cite{Lucca:2019rxf}
\begin{equation}
    \left.\frac{\mathrm{d}^2 E}{\mathrm{d}V \, \mathrm{d}t}\right|_\text{inj,em} = \frac{f_\text{dec}}{\tau_\chi} \zeta_\text{em}(M_\chi) \rho_\text{DM,0} (1+z)^3 e^{-t/\tau_\chi} \; ,
\end{equation}
where we set $\zeta_\text{em} = 1$ for decays into photons and electron-positron pairs. This formula suggests that to first approximation cosmological observables are sensitive only to the product $f_\text{em} = f_\text{decay} \zeta_\text{em}$, which has been exploited in ref.~\cite{Hambye:2021moy} to derive constraints on DM decays into neutrinos. However, to understand in detail the effect of this energy injection on the IGM, it is crucial to distinguish between the injected and the deposited energy. To first approximation, the decay of a TeV-scale particle leads to a small number of highly-energetic particles, which cannot efficiently transfer their energy to heat or ionize the neutral hydrogen gas. It therefore becomes necessary to model the energy loss of these particles through interactions and redshifting using appropriate transfer functions. These transfer functions can be convolved with the injection rate to obtain so-called efficiency functions $f_\text{c}(z, x_e)$, which depend on the channel $c$ (photons, electrons or positrons), the redshift $z$ and the ionization fraction $x_e$, such that the energy deposition can be written as
\begin{equation}
    \left.\frac{\mathrm{d}^2 E}{\mathrm{d}V \, \mathrm{d}t}\right|_\text{dep,c}(z, x_e) = f_c(z, x_e) \cdot \left.\frac{\mathrm{d}^2 E}{\mathrm{d}V \, \mathrm{d}t}\right|_\text{inj,em}(z) \; .
\end{equation}
These efficiency functions have been implemented in \texttt{DarkHistory} in order to calculate the evolution of $x_e$ depending on the model parameters $f_\text{decay}$, $\tau_\chi$, $M_\chi$ and the decay channel.

\section{CMB constraints}
\label{sec:CMB}

To study the effect of exotic energy injection on the CMB, we make use of the on-the-spot approximation, which assumes that energy deposition happens very quickly after injection and exhibits no strong redshift dependence. In this case, the efficiency function can be approximated as~\cite{Poulin:2016anj}
\begin{equation}
f_c(z, x_e) = \chi_c(x_e) f_\text{eff} e^{-t/\tau_\chi} \; ,
\end{equation}
where the functions $\chi_c(x_e)$ can be calculated in advance in a model-independent way and the effective efficiency factor $f_\text{eff}$ depends only on the decay channel and (potentially) on the DM mass.\footnote{We note that some studies have also considered a different approximation, which makes use of the factorisation $f_c(z, x_e) = \chi_c(x_e) f_c(z)$ but allows for a redshift dependence of $f_c(z)$~\cite{Stocker:2018avm,GAMBITCosmologyWorkgroup:2020htv}. However, this development has not been kept up-to-date with subsequent improvements of \texttt{CLASS}. In particular, we assume that some (constant) fraction $f_\text{eff}$ of the electromagnetic energy is instantly absorbed by the medium, and do not treat redshifting and delayed deposition of energy from the decay products. We use this approximation only in our CMB analysis. %\textcolor{red}{However, to the best of our knowledge the functions $f_c(z)$ are not readily available for decays into neutrinos.}
}

\begin{figure}
    \centering
    \includegraphics[width=\linewidth]{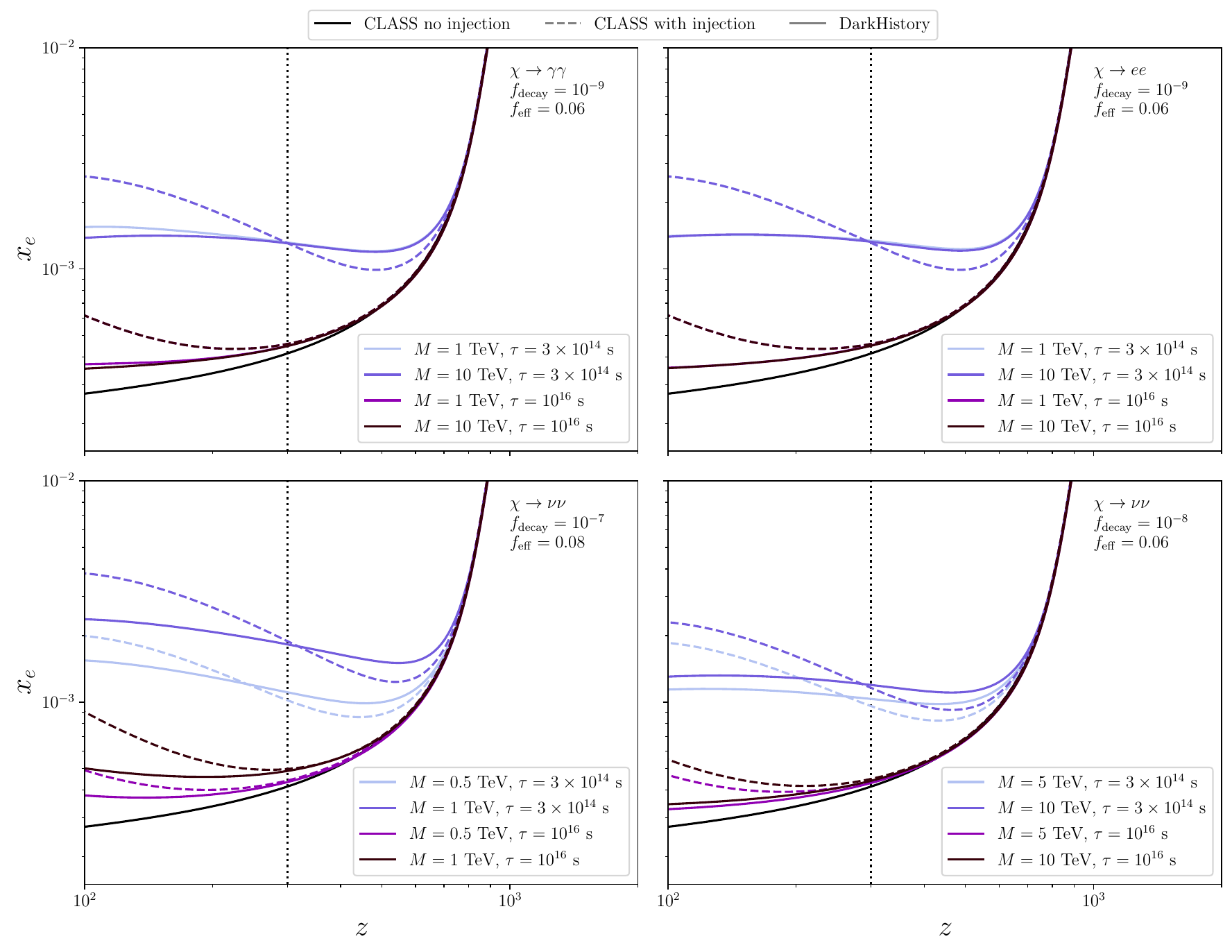}
    \caption{The free electron fraction $x_e$ in decaying DM scenarios computed with \texttt{DarkHistory} (solid colored), \CLASS{} (dashed colored), with the values computed by \CLASS{} for no injection for comparison (solid black). Top left panel shows decays to photons, top right decays to electron-positron pairs, and the bottom panels show decays to neutrinos from low or high DM masses respectively. In the upper panels, the dashed lines for 1 and 10 TeV overlap exactly for each lifetime.}
    \label{fig:efficiencies}
\end{figure}

It has been argued in Refs.~\cite{Slatyer:2016qyl,Poulin:2016anj} that in order to accurately capture CMB constraints, $f_\text{eff}$ should be chosen in such a way that it reproduces the full efficiency function $f_c(z,x_e)$ at redshift $z = 300$. We implement this prescription by determining the value of $f_\text{eff}$ such that the free electron fraction $x_e$ computed by \CLASS{}~\cite{Blas:2011rf,Lesgourgues:2011rh,Stocker:2018avm,Lucca:2019rxf} using the on-the-spot approximation matches the result from \texttt{DarkHistory}. We find that decays to electron-positron pairs and photons are both well-described, (according to the $x_e (z=300)$ criterion), by an efficiency of $f_\text{eff}=0.06$ for the entire range of masses and lifetimes, while the efficiency of decays to neutrino varies between $f_\text{eff}=0.08$ (after accounting for $\zeta_\text{em}$) for the smallest DM masses that we consider and $f_\text{eff}=0.06$ for the  largest. This is illustrated in figure~\ref{fig:efficiencies}.

\begin{figure}
    \centering
    \includegraphics[width=\linewidth]{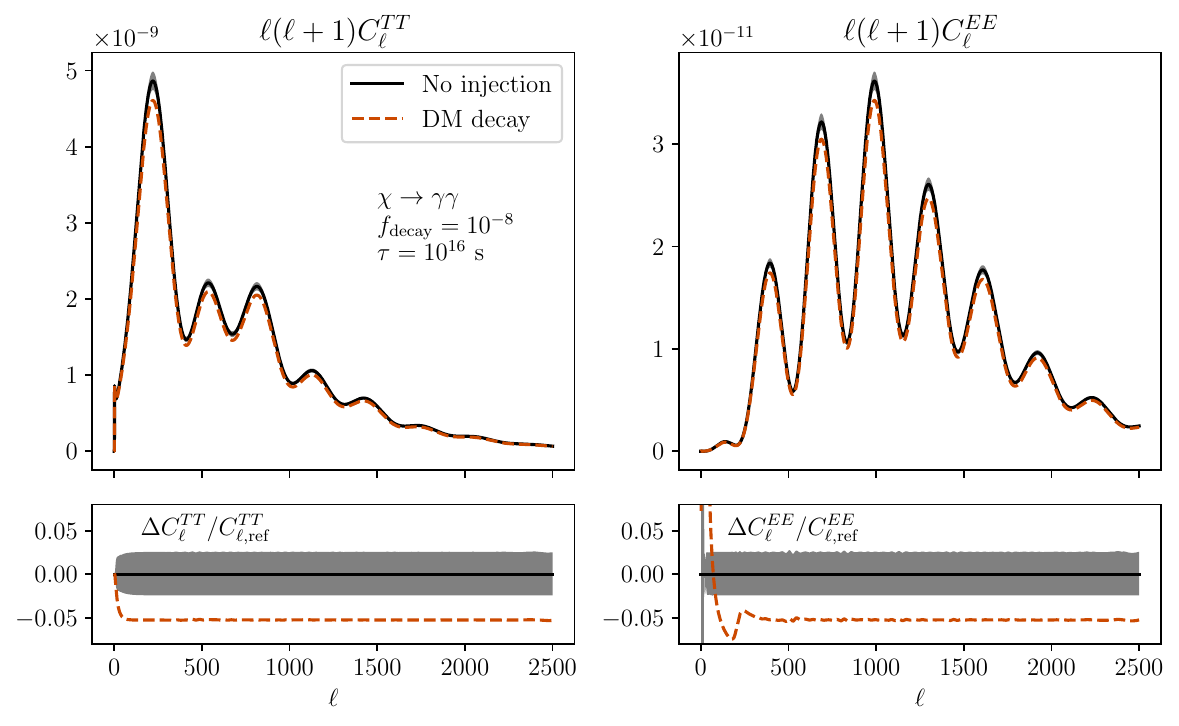}
    \caption{The temperature and polarization power spectra of the CMB for an example model of decaying DM, $\chi \rightarrow \gamma\gamma$, with a lifetime $\tau=10^{16}$ s, and a decaying fraction of $f_\mathrm{decay}=10^{-8}$ (orange dashed), compared to a model with no exotic energy injection (black solid). The grey shaded region corresponds to the 95\%-confidence upper and lower limits on $z_\text{reio}$ obtained from our MCMC. This model can be confidently ruled out, as shown in figure~\ref{fig:CMBlims}.}
    \label{fig:CMB_spectra}
\end{figure}

The main impact on the CMB from the decay of long-lived particles like the ones we investigate in this work comes from their ionizing effect on the IGM, which increases the rate of scattering of the CMB photons, causing an overall suppression of the angular power spectrum of the CMB anisotropies. We show the impact on the CMB multipoles of an example model of decaying DM, $\chi \rightarrow \gamma\gamma$ with a lifetime $\tau=10^{16}$ s and a decaying fraction of $f_\mathrm{decay}=10^{-8}$, in figure~\ref{fig:CMB_spectra}.

We obtain CMB limits on DM decay by performing an MCMC analysis with \CLASS{} and \MontePython{}~\cite{Audren:2012wb,Brinckmann:2018cvx}, using the energy deposition prescription by Galli et al. 2013~\cite{Galli:2013dna} included with the public version of \CLASS{}.
We perform our MCMC with the CMB likelihoods \texttt{CamSpec PR4\_12.7cl}~\cite{Rosenberg:2022sdy}, \texttt{SPT-3G D1 T\&E lite} (\texttt{candl})~\cite{SPT-3G:2025bzu,Balkenhol:2024sbv}, \texttt{ACT DR6 TT/TE/EE lite} (\texttt{candl})~\cite{AtacamaCosmologyTelescope:2025vnj,AtacamaCosmologyTelescope:2025blo,AtacamaCosmologyTelescope:2025nti}, \texttt{ACT DR6 lensing}~\cite{ACT:2023kun,ACT:2023dou}, \texttt{SPT-3G 2yr lensing} (\texttt{MUSE})~\cite{SPT-3G:2024atg}, \texttt{Planck 2018 low\_l TT}~\cite{Planck:2019nip}, and \texttt{Planck SRoll2 low\_l EE}~\cite{Delouis:2019bub}.
We set our limit at the 95\% upper confidence limit on the decaying fraction for each decay width $\Gamma_\mathrm{decay}$, with our chains converged to a Gelman-Rubin statistic level of $R-1<0.01$.

\begin{figure}
    \centering
    \includegraphics[width=0.6\linewidth]{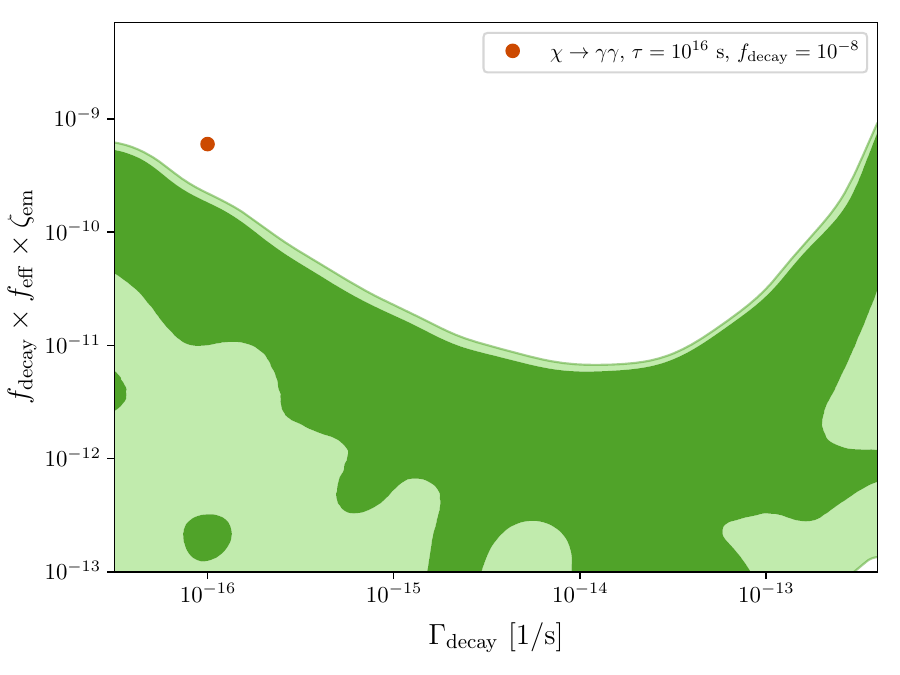}
    \caption{Two-dimensional posterior distribution of the decay parameters $f_\mathrm{decay}$ and $\Gamma_\mathrm{decay}$, obtained with \CLASS{} and \MontePython{}. The orange circle shows the point in parameter space illustrated in figure~\ref{fig:CMB_spectra}. Figure produced using GetDist~\cite{Lewis:2019xzd}.}
    \label{fig:CMBlims}
\end{figure}

We perform our MCMC using a standard set of six cosmological $\Lambda$CDM parameters $\left\{ \omega_\mathrm{b},  \omega_\mathrm{cdm}, H_0, \ln\left( 10^{10} A_\mathrm{s}\right), n_\mathrm{s}, z_\mathrm{reio} \right\}$, plus neutrino masses (assuming three degenerate mass states) $m_\nu$, and the decay parameters $\left\{ \log_{10}\left( f_\mathrm{decay} \right), \log_{10}\left( \Gamma_\mathrm{decay} \right)\right\} $. A wide, flat prior was used for all parameters except those governing the decays. We put a prior on the decay width $\log_{10}\left( \Gamma_\mathrm{decay}\right) \in \left[-16.5, -12.4\right]$ to restrict the space of lifetimes probed to those most relevant, and a prior $ \log_{10}\left( f_\mathrm{decay} \right) \in \left[ -13, -6 \right] $ (well below sensitivity and strongly ruled out, respectively) to speed up convergence by preventing the chain from exploring infinitely negative values. Both of these priors were likewise flat in the probed log-space. Under the on-the-spot approximation used, $f_\mathrm{decay}$, $f_\mathrm{eff}$ and $\zeta_\text{em}$ are completely degenerate (the relevant quantity being $f_\mathrm{decay} \times f_\mathrm{eff} \times \zeta_\text{em}$), we therefore include the efficiency by rescaling the limits by a factor of $f_\mathrm{eff}$ post facto. Our limits are shown in figure~\ref{fig:CMBlims}. The dark (light) green shaded region corresponds to the credible parameter region at 68\% (95\%) probability, while its upper boundary can be approximately interpreted as the 95\% confidence level upper bound on $f_\text{decay} \times f_\text{eff} \times \zeta_\text{em}$. The fact that the posterior peaks at non-zero values of $f$ is a numerical artifact of the parameter scan. The true posterior becomes flat for very small values of $f$. After accounting for the effect of $f_\text{eff}$ on the bounds, our CMB bounds are consistent with those obtained in ref.~\cite{Lopez-Honorez:2026bzj} using only the Planck bound on $\tau_\text{reio}$ rather than a full MCMC pipeline, but with a more sophisticated treatment of the energy deposition from DM decay. The main weakness, shared by both analyses, is the simplistic $\tanh$-parameterization of reionisation, which does not treat the non-linear nature of reionisation in a fully physical and self-consistent way. However, extending \CLASS{} with such features is beyond the scope of this work. Importantly, our 21-cm analysis, which is much more sensitive to this, relies on the non-linear treatment of \texttt{21cmFAST}, which we will now describe in detail.

\section{Impact of exotic energy injection on the 21-cm signal}
\label{sec:21cm}

We compute the impact of DM decays on the global 21-cm signal using \texttt{DM21cm}~\cite{Sun:2023acy}, which relies on \texttt{DarkHistory}~\cite{Liu:2019bbm,Sun:2022djj} to compute the energy injection history, and \texttt{21cmFAST}~\cite{Mesinger:2010ne,Murray:2020trn,Park:2018ljd,Qin:2020xyh,Munoz:2021psm} to evolve the dynamics of the neutral hydrogen. We use a slightly modified version of the decay treatment available in the public version of \texttt{DM21cm} to incorporate the depletion of decaying DM with lifetimes significantly shorter than the age of the universe, such as those we consider in this work. We give here a very brief overview of the relevant 21-cm physics. For more details we refer to the papers describing \texttt{DM21cm}~\cite{Sun:2023acy} and \texttt{21cmFAST}~\cite{Mesinger:2010ne,Greig:2018hja,Park:2018ljd,Murray:2020trn,Qin:2020xyh,Munoz:2021psm}.

The observable in 21-cm cosmology is the brightness temperature of the 21-cm hyperfine transition line of neutral hydrogen, measured along a line of sight and compared to the blackbody spectrum of the CMB. This temperature is frequency-dependent approximately given as~\cite{Sun:2023acy}
\begin{equation}
    T_{21} (\nu) \approx 27 h_\mathrm{HI} \left( 1+\delta \right) \left( \frac{H}{dv_r/dr + H} \right) \left( 1- \frac{T_\text{cmb}}{T_S} \right) \left( \frac{1+z}{10} \frac{0.15}{\omega_\text{m}} \right)^{1/2}\left( \frac{\omega_\text{b}}{0.023} \right) \, \text{mK} \; ,
\end{equation}
where $\delta$ is the Eulerian density contrast, $T_\text{cmb}$ and $T_S$ are the temperatures of the CMB and the atomic hydrogen spin, respectively, $\omega_\text{m}$ and $\omega_\text{b}$ are the present day physical matter and baryon energy densities, $H$ is the Hubble parameter, and $dv_r/dr$ is the comoving gradient of the comoving line-of-sight velocity of the gas. %\textcolor{red}{$h_\mathrm{HI}$?}

Arguably the most important -- and involved -- parameter is the neutral hydrogen spin temperature $T_S$, which parametrises the relative occupation of the two spin states  of the hyperfine splitting in the atomic hydrogen ground state,
\begin{equation}
    \frac{n_1}{n_0} \equiv 3 \exp \left( -T_* / T_S \right),
\end{equation}
where $T_*$ is the temperature corresponding to the energy split between the two states, $T_* = 0.068\, \text{K} = 5.9\times 10^{-6}\, \text{eV}$. The hyperfine transitions that govern the spin temperature have three contributions: atomic collisions, absorption or emission of CMB photons, and the absorption or emission of Lyman-$\alpha$ photons (the latter is known as the Wouthuysen-Field effect~\cite{1952AJ.....57R..31W,1959ApJ...129..536F}). Accordingly, the spin temperature can be written as,
\begin{equation}
    T_S^{-1} = \frac{T^{-1}_\text{cmb} + x_c T^{-1}_k + x_\alpha T_\alpha^{-1}}{1+x_c+x_\alpha} \; ,
\end{equation}
where $T_k$ and $T_\alpha$ are the kinetic gas temperature and the effective colour temperature of the Lyman-$\alpha$ radiation field, respectively, while $x_c$ and $x_\alpha$ are the respective coupling coefficients for collisions and scattering with Lyman-$\alpha$ photons.

Before reionisation, the kinetic gas temperature is much smaller than the CMB temperature, such that $T_k \leq T_S \leq T_\text{cmb}$. At redshift $z \sim 25\text{--}30$ the couplings $x_c$ and $x_\alpha$ are typically so small that $T_S \approx T_\text{cmb}$ and therefore $T_{21}(\nu) \approx 0$. At redshift $z \sim 15\text{--}20$, the light from the first stars strongly increases the coupling to Lyman-$\alpha$ photons, such that $T_S$ decreases rapidly and approaches $T_b$. This decrease in temperature leads to an absorption of CMB photons, such that $T_{21} < 0$, which ends when reionisation leads to a rapid increase of the temperature of baryonic gas.

The main effect of DM decays (apart from modifying the ionization fraction $x_e$ already discussed above) is to inject heat into the baryonic gas. This heating term leads to a slower decrease of the kinetic gas temperature with decreasing redshift than in standard cosmology. As a result, $T_S$ at redshift $z \approx 15$ is larger than in the absence of DM decays, and the depth of the absorption feature in the 21-cm brightness temperature is reduced. An observation of this feature therefore makes it possible to constrain decaying DM subcomponents.

To calculate the magnitude of this effect, all of the relevant quantities are evolved in a three-dimensional box by \texttt{21cmFAST}, following a set of coupled differential equation, taking into account the inhomogeneous matter field, and its non-linear perturbations. \texttt{DM21cm} interfaces its timestepping with \texttt{DarkHistory}, which computes the injection spectra of the decaying DM species, and how the energy is deposited into both the baryonic matter, and as a background radiation field of photons. The brightness temperature is then averaged across the box to obtain the global signal.

\begin{figure}
    \centering
    \includegraphics[width=\linewidth]{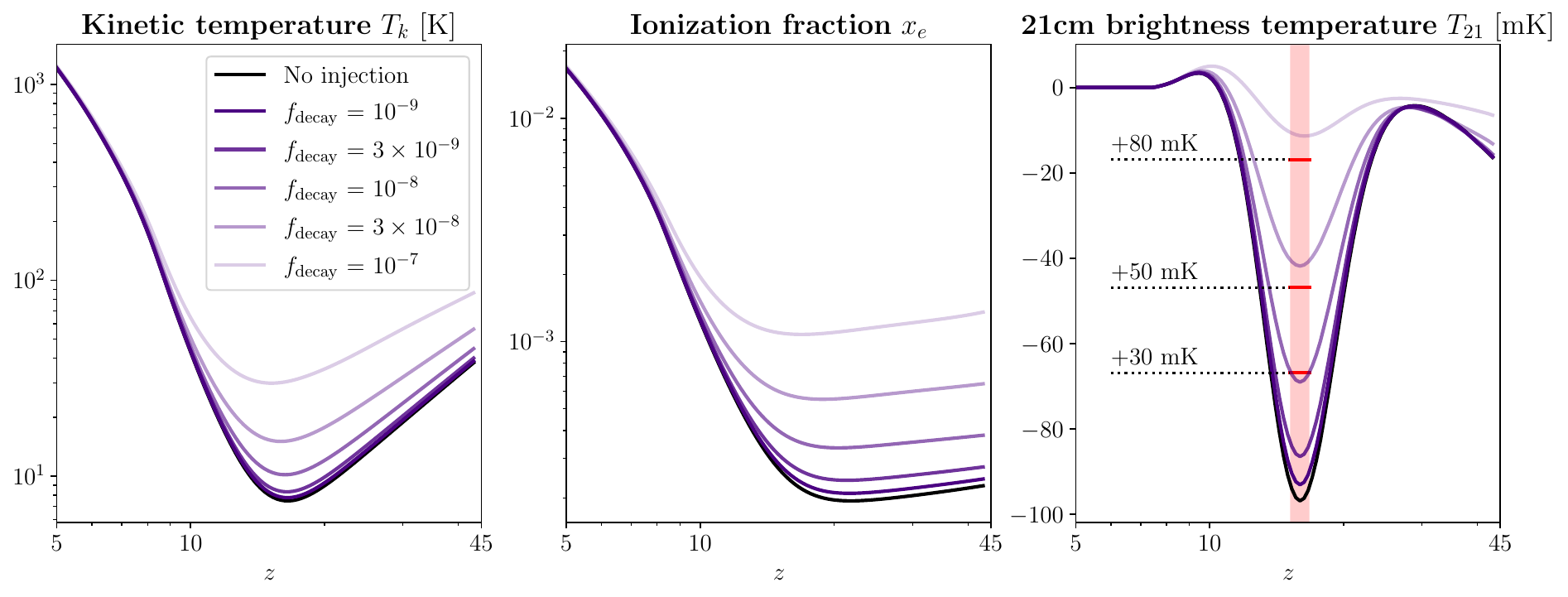}
    \caption{Redshift evolution of kinetic temperature, ionization fraction, and 21-cm temperature for a range of decaying DM fractions, all for decays to neutrinos with lifetime $\tau = 3\times10^{15}$ s and mass 10 TeV.}
    \label{fig:temp_evol}
\end{figure}

We show examples of the evolution of the globally averaged kinetic temperature and ionization fraction, as well as the resulting 21-cm brightness temperature in figure~\ref{fig:temp_evol}. In this figure we consider decays into neutrinos and fix the DM mass to $M_\chi = 10\,\mathrm{TeV}$ and the lifetime to $\tau = 3 \times 10^{15} \, \mathrm{s}$, while the fraction of decaying DM is varied between $f_\text{decay} = 10^{-9}$ and $10^{-7}$. For the smallest value of $f_\text{decay}$ the evolution is virtually indistinguishable from standard cosmology, while for the largest value, the kinetic gas temperature becomes so large that the absorption feature disappears almost entirely. Clearly, such a scenario would be strongly ruled out if an absorption feature is observed.

To derive sensitivity projections for the full parameter space, we evolve three grids of models with decays into photons, neutrinos, and electron-positron pairs respectively, with masses of the decaying DM particle and lifetimes
\begin{align*}
    M_\chi &\in \left[ 0.5, 1, 2, 5, 10 \right] \, \text{TeV} \\
    \tau_\chi &\in \left[ 3\cdot 10^{14}, 10^{15},3\cdot 10^{15}, 10^{16},3\cdot 10^{16} \right] \, \text{s}
\end{align*}
and decaying fractions 
\begin{equation*}
    f_\mathrm{decay} \in \left[  3\cdot 10^{-11}, 10^{-10},3\cdot 10^{-10}, 10^{-9},3\cdot 10^{-9}, 10^{-8},3\cdot 10^{-8}, 10^{-7}\right]
\end{equation*}
for decays to photons or electron-positron pairs, or 
\begin{equation*}
    f_\mathrm{decay} \in \left[  3\cdot 10^{-10}, 10^{-9},3\cdot 10^{-9}, 10^{-8},3\cdot 10^{-8}, 10^{-7},3\cdot 10^{-7}, 10^{-6}\right]
\end{equation*}
for decays to neutrinos. We use a box size of 256 conformal Mpc, and $64^3$ cells for the low-resolution \texttt{21cmFast} box, as we are interested in the global signal, rather than intensity mapping. We keep all other settings and astrophysical parameters at their \texttt{DM21cm} defaults, for a full list, see appendix~\ref{app:astro_params}.

\begin{figure}
    \centering
    \includegraphics[width=0.8\linewidth]{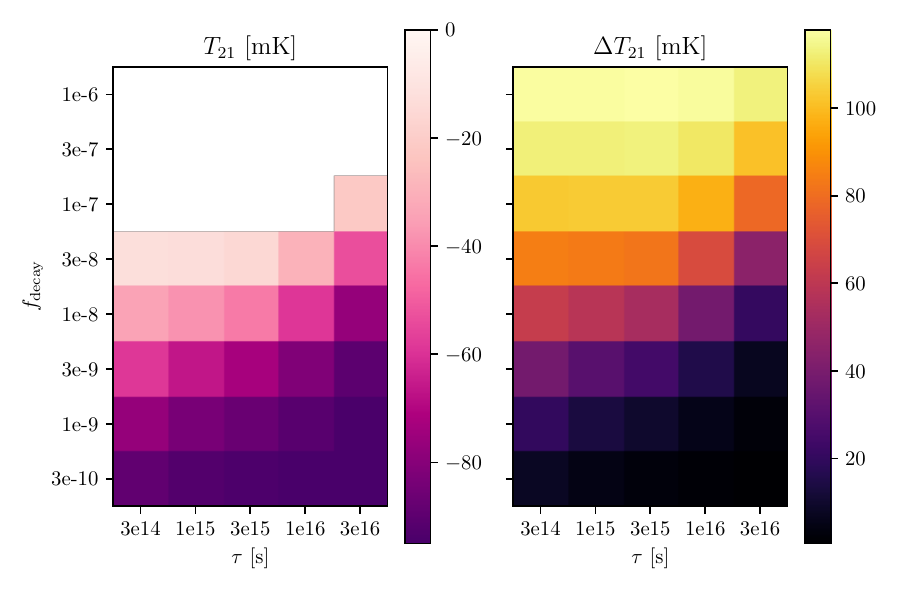}
    \caption{Impact on the 21-cm brightness temperature from a DM particle with a mass of 10 TeV decaying to neutrinos. \textbf{Left:} Absolute temperature of the global 21-cm signal at its $z\sim16$ minimum. White squares indicate a temperature that is no longer negative. \textbf{Right:} The difference in global  21-cm temperature at the $z\sim16$ minimum, relatively to a model with no exotic energy injection.}
    \label{fig:tempgrid}
\end{figure}

As a proxy for the detectability of the energy injection, we use the change to the global 21-cm temperature, $\Delta T_{21}$ at its minimum around $z\sim16$ with respect to a fiducial $\Lambda$CDM model with no exotic energy injection. In figure~\ref{fig:tempgrid}, we show both $T_{21}$ and $\Delta T_{21}$ as a function of $f_\text{decay}$ and $\tau$ for $M_\chi = 10 \, \mathrm{TeV}$ and decays into neutrinos. We see that the effect is largest for $\tau \leq 3 \times 10^{15} \, \mathrm{s}$ corresponding to the case where most of the injection happens above redshift $z \sim 20$.

Based on these results, it is in principle possible to calculate the projected sensitivity of a given instrument. Doing so would however also require a variation of the astrophysical parameters that enter the simulations. 
For example,  ref.~\cite{Facchinetti:2023slb} has studied the impact of astrophysical uncertainties for a specific instrument by performing a Fisher analysis, and ref.~\cite{Agius:2025xbj} has shown that the energy injected into the medium from astrophysical sources can have significant impact on the extracted limits of primordial black holes, effects which were also pointed out in ref.~\cite{Lopez-Honorez:2013cua}.
Here we take a simpler approach and adopt three hypothetical thresholds for detection at $\Delta T_{21}= 30$ mK, $50$ mK, and $80$ mK, representing an optimistic, intermediate, and pessimistic assumption for the experimental sensitivity, respectively. For each threshold, we then determine the minimal decay fraction that may be observable for each combination of DM mass, lifetime, and decay products. 

\section{Results}
\label{sec:results}

\begin{figure}
    \centering
    \includegraphics[width=\linewidth]{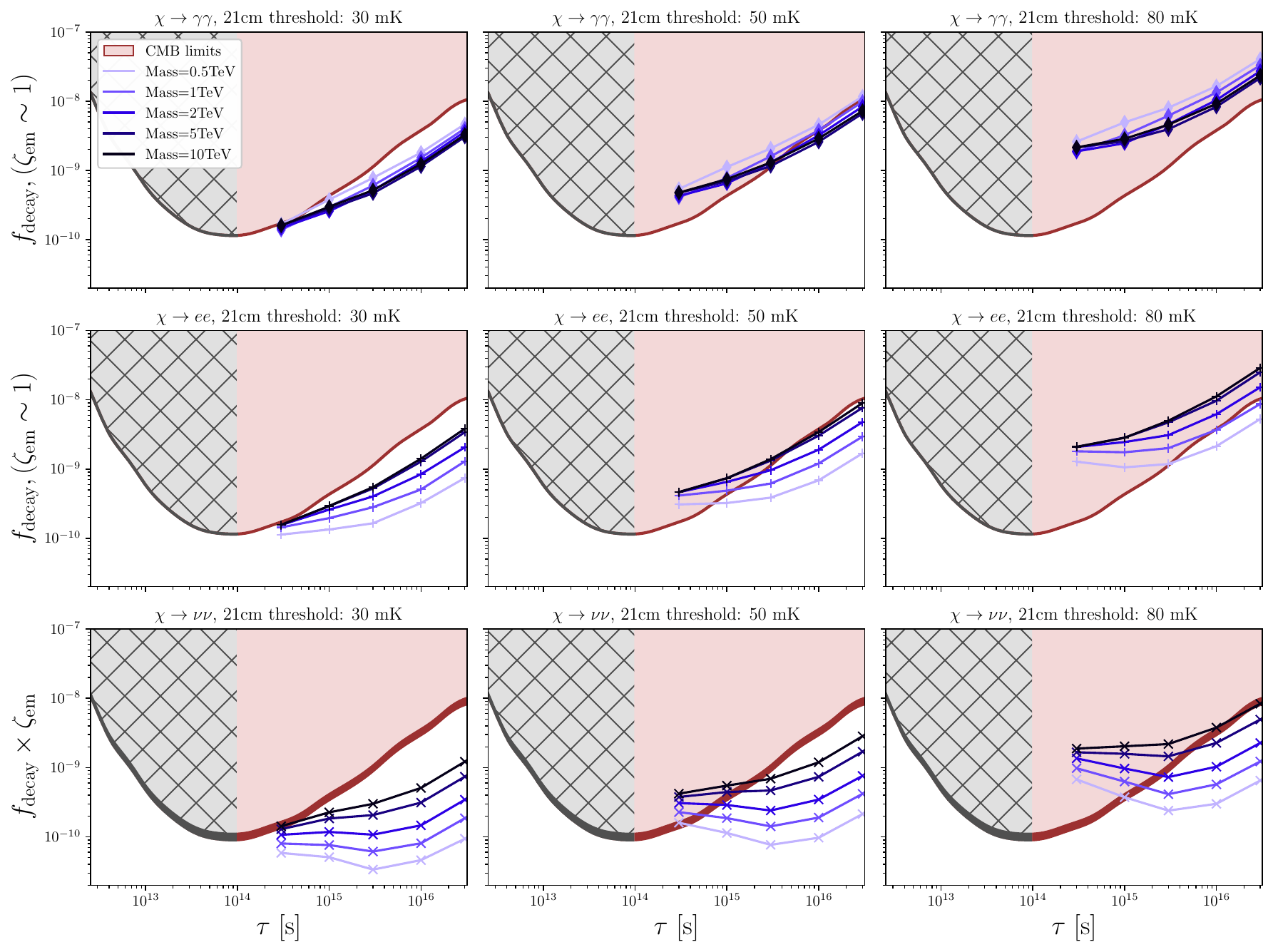}
    \caption{Projected constraints on energy injection from the 21-cm global temperature as well our newly derived CMB constraints. The grey hatched region $\tau < 10^{14}$ s indicates that the on-the-spot approximation is expected to be less reliable for such short lifetimes.}
    \label{fig:constraintgrid}
\end{figure}

A collection of all the 21-cm sensitivity projections in terms of $\tau$ and $f_\text{decay}$ compared to the CMB bound is shown in figure~\ref{fig:constraintgrid}. Different rows correspond to different final states ($\gamma\gamma$, $e^+ e^-$ and $\nu \nu$ from top to bottom), while different columns correspond to the optimistic, intermediate and pessimistic assumption for the 21-cm sensitivity (see above). In each panel the blue lines show the 21-cm sensitivities for different DM masses, while the red line corresponds to the CMB bound (with the width of the line reflecting the change in $f_\text{eff}$ with varying DM mass).

For decays into photons, there is almost no dependence of the sensitivity projections on the mass of the decaying DM particle. This is true also for the case of DM annihilations~\cite{Slatyer:2015jla} and consistent with the fact that the energy deposition efficiency $f(z)$ exhibits no sharp features as a function of the photon mass~\cite{Slatyer:2015kla}. The projected bounds generally become weaker with increasing DM lifetime and surpass the existing CMB limits only in the optimistic case. For decays into electrons the dependence on the DM mass is more pronounced, with smaller masses corresponding to better sensitivities. This is because lower-energy electrons require less energy loss into photons via inverse Compton scattering before they reach the MeV energy range, where they become very efficient in depositing their energy through collisional processes with the  gas~\cite{Slatyer:2015kla}. As a result, the expected bounds from a 21-cm signal become relevant also in the intermediate case, where we require a temperature difference of 50 mK.

The dependence on the DM mass is even more enhanced for neutrinos. If we account for the fraction of the decay mass going into electromagnetic energy (i.e.\ we place bounds on the combination $f_\text{decay} \times \zeta_\text{em}(M_\chi)$), we find that the sensitivity of future 21-cm signals is substantially enhanced for the smallest DM masses that we consider. The reason is that the injection of electromagnetic energy from decays into neutrinos happens dominantly at lower energies.
For example, for $m_\chi = 1 \, \mathrm{TeV}$ and decays into electron-positron pairs, around 90\% of the injected electromagnetic energy will be in the form of electrons and positrons with an energy of more than 300 GeV, while the fraction is only 30\% for decays into neutrinos. Since lower-energy electrons deposit their energy more efficiently, decays into neutrinos therefore have a larger effect on the 21-cm signal than decays into photons or electrons if the total amount of injected electromagnetic energy is fixed. Of course, since $\zeta_\text{em}(M_\chi) \ll 1$, the direct bound on $f_\text{decay}$ is still stronger for decays into electrons and photons, and becomes stronger with increasing DM mass (and hence increasing $\zeta(M_\chi)$) for decays into neutrinos (see appendix~\ref{app:nu_decays}). 

As discussed in section~\ref{sec:CMB}, the CMB is sensitive mostly to the energy deposition at redshift $z \sim 300$, which depends almost exclusively on the total amount of electromagnetic energy injected (with $f_\text{eff}$ varying only between 0.06 and 0.08 with varying DM mass and decay channel). For the 21-cm signal, on the other hand, the energy deposition at much later times becomes relevant, which is much more sensitive to the detailed distribution of the injected electromagnetic energy. We therefore conclude that the 21-cm signal is particularly sensitive to the case of DM decays into neutrinos, for which the effects at late times are enhanced relative to the effects at early times. This means that even in the most pessimistic case the imprint of DM decays on the 21-cm signal may be seen in regions of parameter space compatible with the CMB bound.

The differences between the different decay channels is most pronounced for DM masses at or below 1 TeV. At higher masses, the expectation is that the effect of decays into photons and electrons should be the same, since both particle species induce an electromagnetic cascade via pair production and inverse Compton scattering on CMB photons. Moreover, the injection spectrum of electromagnetic energy for decays into neutrinos becomes more and more similar to the one for electrons and photons. Already for $M_\chi = 10 \, \mathrm{TeV}$ more than 60\% of the injected electromagnetic energy is in the form of electrons and positrons with an energy of more than 3 TeV. 

\begin{figure}
    \centering
    \includegraphics[width=\linewidth]{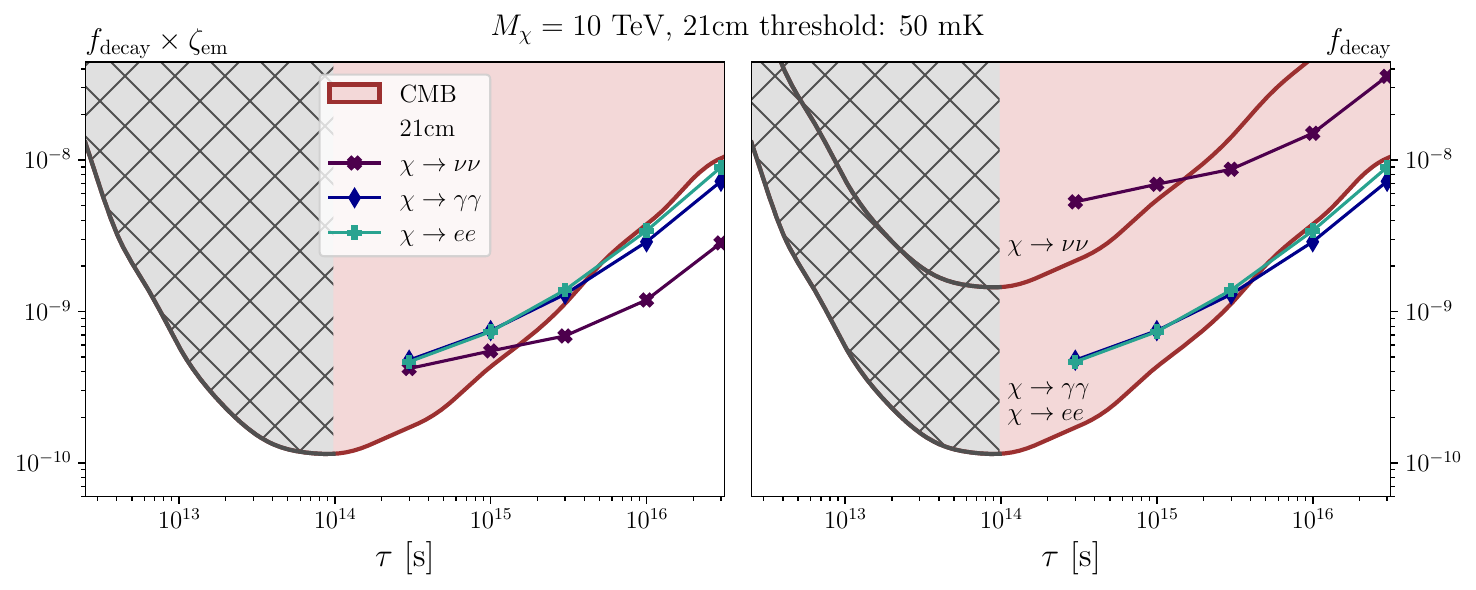}
    \caption{Limits on the decaying fraction for a 10 TeV decaying particle and a $\Delta T_{21}=50$ mK detection threshold for all three decay channels. The grey hatched region $\tau < 10^{14}$ s indicates that the on-the-spot approximation is expected to be less reliable for such short lifetimes.}
    \label{fig:10TeV_50mK_all}
\end{figure}

We confirm this expectation in figure~\ref{fig:10TeV_50mK_all} where we show the bounds for decays into photons, electrons and neutrinos for $M_\chi = 10 \, \mathrm {TeV}$ in the same plot. In the left panel, we show the bounds on $f_\text{decay} \times \zeta_\text{em}(M_\chi)$, while the right panel shows just the bound on $f_\text{decay}$, such that both the CMB bound and the 21-cm sensitivity are weaker for decays into neutrinos by approximately an order of magnitude. For the smallest lifetimes that we consider ($\tau = 3 \cdot 10^{14} \, \mathrm{s}$) we see that indeed the 21-cm signal depends only on the total amount of electromagnetic energy injected, not on its spectral distribution. For longer lifetimes, the sensitivity for decays into neutrinos is again somewhat better than for decays into electrons or photons. However, we expect that for DM masses above 10 TeV (which are currently not supported by \texttt{DarkHistory} due to the lack of interpolation tables) the 21-cm sensitivity for decays into neutrinos will converge towards the lines for decays into electrons and photons, because the electromagnetic cascade will be very similar in all cases. Our results thus suggest that a conservative estimate of the 21-cm sensitivity for $M_\chi > 10 \, \mathrm{TeV}$ can be obtained for any annihilation channel by simply considering the total amount of injected electromagnetic energy and comparing to the case of decays into photons or electrons.

While a full scan of astrophysical parameters across the range of models is beyond the scope of this work, we investigate the impact of a few that are most likely to affect the sensitivity to energy injection from dark matter decay. We choose the parameters \texttt{F\_STAR7\_MINI}, \texttt{F\_STAR10}, \texttt{M\_TURN}, and, \texttt{L\_X}. The two \texttt{F\_STAR} parameters set the fraction of galactic gas in stars in minihaloes and normal haloes respectively, as described in Refs.~\cite{Greig:2018hja,Park:2018ljd,Qin:2020xyh}. \texttt{M\_TURN} is the turnover mass, setting the scale for quenching of star formation due to supernova and photo-heating feedback, as described in Ref.~\cite{Park:2018ljd}. \texttt{L\_X} is the specific X-ray luminosity escaping host galaxies per unit of star formation as described in Refs.~\cite{Greig:2018hja,Qin:2020xyh}. \texttt{L\_X\_MINI} sets the corresponding value for minihaloes, which we keep equal to \texttt{L\_X}. All these parameters are given in $\log_{10}$ units, so varying each by 1 unit corresponds to an order of magnitude change in the astrophysical parameter.

\begin{figure}[t]
    \centering
    \includegraphics[width=\linewidth]{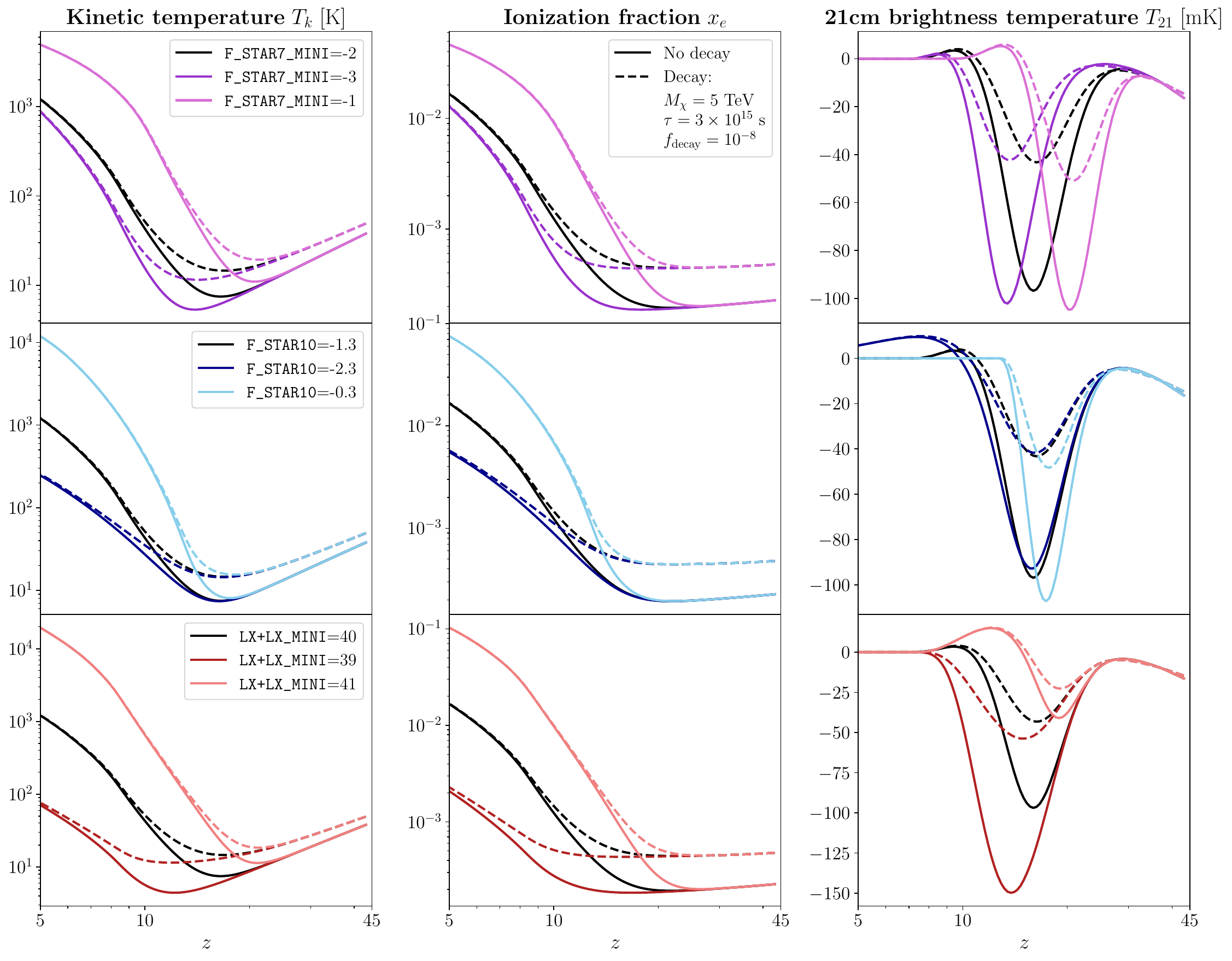}
    \caption{Redshift evolution of kinetic temperature, ionization fraction, and 21-cm temperature for different values of the astrophysical parameters, showing both cosmologies with no energy injection from DM decay (solid), and with a model with DM decaying to neutrinos (dashed) with DM mass $M_\chi=5$ TeV, lifetime $\tau=3\times 10^{15}$ s, and a decaying fraction $f_\text{decay}=10^{-8}$.}
    \label{fig:temp_evol_astro}
\end{figure}

We find that the effect of changing \texttt{M\_TURN} is entirely negligible, and show the rest of the parameters in Fig.~\ref{fig:temp_evol_astro}. From the results it is clear that the effects of the \texttt{F\_STAR} parameters are very distinct from the impact of DM decay, primarily affecting the redshift at which the 21cm brightness temperature is at its minimum, without greatly affecting the exact minimum value. The \texttt{L\_X}) parameter is a bit more problematic, shifting both the redshift and depth of the minimum. The case of an increased X-ray luminosity has a depth of the minimum comparable to the injection model with fiducial astrophysical parameters, however, it also has a very characteristic enhancement of the positive bump at slightly lower redshifts, so with realistic observations measuring more than just the depth of the minimum, an energy injection from DM decay would be distinguishable from changes in the tested astrophysical parameters with the sufficient experimental sensitivity. %a large value of the X-ray luminosity. 
It is also worth noting that our variations in the \texttt{L\_X}) parameter are very extreme, changing it by $\pm1$ from the fiducial value of 40, where Ref.~\cite{Greig:2018hja} found that it could be constrained at the level of $\pm0.05$. However, the decreased depth of the minimum does limit the possible space for decaying DM to impact observables, so while the effects are not degenerate, limits on energy injection from the dark sector will be weaker, the larger the energy injection from astrophysical sources.

\section{Conclusions}
\label{sec:conclusion}

In this work, we have investigated the impact on the early universe of dark matter subcomponents decaying between recombination and reionisation into either photons, electron-positron pairs or neutrinos. For this purpose we have performed simulations using the \texttt{DM21cm} code~\cite{Sun:2023acy}, which relies on \texttt{21cmFAST}~\cite{Mesinger:2010ne,Greig:2018hja,Park:2018ljd,Murray:2020trn,Qin:2020xyh,Munoz:2021psm} for its evolution of the intergalactic medium, radiation field, and the 21-cm temperature, and \texttt{DarkHistory}~\cite{Liu:2019bbm,Sun:2022djj} for the computation of dark matter decays and the energy injection parameters. We have then obtained sensitivity forecasts based on the deviation of the 21-cm brightness temperature from the standard expectation at its $z\sim 16$ minimum, with optimistic, intermediate, and pessimistic detection thresholds at 30, 50, and 80 mK. We have also presented CMB limits for the same dark matter models based on a full MCMC analysis using state-of-the-art CMB datasets, showing good agreement with those from~ref.~\cite{Lopez-Honorez:2026bzj}.

Our simulation results show that a precision measurement of the global 21-cm temperature signal can provide a strong complement to the CMB for probing dark matter particles that decay between recombination and reionisation. We demonstrate that especially for the case of decays to neutrinos, the 21-cm signal has the potential to substantially improve the CMB limits on the fraction of decaying dark matter for dark matter lifetimes $\tau\gtrsim10^{15}$ s. Even for the pessimistic detection threshold, we find potential to improve upon the CMB limits for decays into neutrinos of dark matter particles with a mass around 1 TeV. For heavier dark matter masses, the differences between photon, electron and neutrino final states become less pronounced (after correcting for the different fractions of electromagnetic energy), and it becomes harder for the 21-cm forecasts to compete with CMB limits. 

Varying the most relevant astrophysical parameters of our 21cm analysis, we find that the effect of energy injection from DM decay is generally distinguishable from astrophysical effects, but scenarios with large energy injection from astrophysical sources, such as a high escaping X-ray luminosity would make the impact of exotic energy injection harder to detect, and thus weaken the potential limits.

Various directions for future research remain. First, it will be interesting to  perform a deeper and more quantitative assessment of the sensitivity of our results to uncertainties in the astrophysics of early star formation, for example using a Fisher matrix analysis as in~\cite{Facchinetti:2023slb}, and more advanced proxies for detectability. Furthermore, unlike our 21-cm analysis, the CMB bounds rely on a rather naive treatment of reionisation, and improving this is an important goal for future research. Finally, the available tabulations limit our analysis to dark matter masses below 10 TeV. While we have presented a plausible way to extrapolate our results to larger masses based on the total amount of injected electromagnetic energy, it would be preferable to extend the underlying codes to study such scenarios explicitly. Together these improvements will provide a more comprehensive and detailed picture of the properties of dark matter and its possible effects on the Dark Ages and the Cosmic Dawn.

\acknowledgments
We would like to thank Laura Lopez-Honorez for helpful comments on the manuscript and Julien Lesgourgues and Tracy Slatyer for useful discussions and feedback. This work has been funded by the DFG through the grant KA 4662/4-1. 

\appendix
\section{Forecast 21-cm limits on decays to neutrinos}
\label{app:nu_decays}

\begin{figure}[b]
    \centering
    \includegraphics[width=\linewidth]{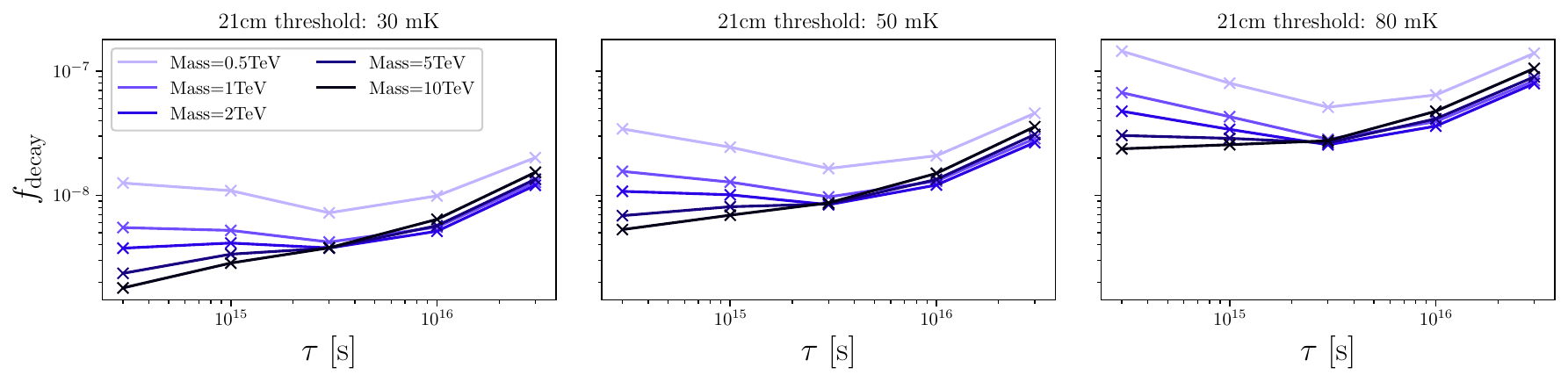}
    \caption{Projected constraints on energy injection from the 21-cm global temperature for DM decaying to neutrinos.}
    \label{fig:neutrinoconstraints}
\end{figure}

We present here for the reader's convenience the full set of limits on DM decaying to netrinos, without the rescaling according to $\zeta_\text{em}$, shown in figure~\ref{fig:neutrinoconstraints}.

\section{Astrophysical parameters in \texttt{DM21cm} simulations}
\label{app:astro_params}
For reference, we include here in Table~\ref{tab:21cm_params} the full list of astrophysics parameters used when performing our 21-cm analysis. See the \texttt{21cmFAST} documentation\footnote{\url{https://21cmfast.readthedocs.io/en/stable/}} for a full description of these and their role.
\begin{table}
    \centering
    \begin{tabular}{|c|c|}
    \hline
         Parameter & Value \\
         \hline
         \texttt{USE\_HALO\_FIELD}& \texttt{False}\\
\texttt{USE\_MINI\_HALOS}& \texttt{True}\\
\texttt{USE\_MASS\_DEPENDENT\_ZETA}& \texttt{True}\\
\texttt{SUBCELL\_RSD}& \texttt{False}\\
\texttt{INHOMO\_RECO}& \texttt{True}\\
\texttt{USE\_TS\_FLUCT}& \texttt{True}\\
\texttt{M\_MIN\_in\_Mass}& \texttt{False}\\
\texttt{PHOTON\_CONS}& \texttt{False}\\
\texttt{FIX\_VCB\_AVG}& \texttt{False}\\
\texttt{HII\_EFF\_FACTOR}& 30.0\\
\texttt{F\_STAR10}& -1.3\\
\texttt{F\_STAR7\_MINI}& -2.0\\
\texttt{ALPHA\_STAR}& 0.5\\
\texttt{ALPHA\_STAR\_MINI}& 0.5\\
\texttt{F\_ESC10}& -1.0\\
\texttt{F\_ESC7\_MINI}& -2.0\\
\texttt{ALPHA\_ESC}& -0.5\\
\texttt{M\_TURN}& 8.7\\
\texttt{R\_BUBBLE\_MAX}& \texttt{None}\\
\texttt{ION\_Tvir\_MIN}& 4.69897\\
\texttt{L\_X}& 40.0\\
\texttt{L\_X\_MINI}& 40.0\\
\texttt{NU\_X\_THRESH}& 500.0\\
\texttt{X\_RAY\_SPEC\_INDEX}& 1.0\\
\texttt{X\_RAY\_Tvir\_MIN}& \texttt{None}\\
\texttt{F\_H2\_SHIELD}& 0.0\\
\texttt{t\_STAR}& 0.5\\
\texttt{N\_RSD\_STEPS}& 20\\
\texttt{A\_LW}& 2.00\\
\texttt{BETA\_LW}& 0.6\\
\texttt{A\_VCB}& 1.0\\
\texttt{BETA\_VCB}& 1.8\\
\hline
    \end{tabular}
    \caption{Full list of astro parameters for our 21cm analysis.}
    \label{tab:21cm_params}
\end{table}

\bibliography{bibliography}

\end{document}